\documentclass[aps,12pts,prl,amsmath,amssymb,preprint,showkeys]{revtex4}
\usepackage{graphicx}
\usepackage{bm}
\usepackage{subfigure}
\usepackage{mathptmx}
\usepackage{epstopdf}
\usepackage[usenames,dvipsnames]{color}

\begin{document}
\title{Ultrafast switching of the electric polarization and magnetic chirality \\ in BiFeO$_3$ by an electric field }
\author{Satadeep Bhattacharjee$^{1}$, Dovran Rahmedov$^{1}$, Dawei Wang$^{2}$, Jorge \'I\~niguez$^{3}$  and Laurent Bellaiche$^{1}$}

\affiliation{$^{1}$Department of Physics and Institute for Nanoscience and Engineering, 
University of Arkansas\\ Fayetteville, Arkansas 72701, USA \\
    $^{2}$Electronic Materials Research Laboratory, Key Laboratory of the Ministry of Education and International Center for Dielectric
  Research, Xi'an Jiaotong University, Xi'an 710049, China \\
  $^{3}$Institut de Ci\`encia de Materials de Barcelona (ICMAB-CSIC), 
Campus UAB, 08193 Bellaterra, Spain}

\begin{abstract}
Using a first-principles-based effective Hamiltonians within molecular dynamics simulations, 
we discover that applying an electric field that is opposite to the initial direction of the polarization results in a switching of both the polarization and the magnetic chirality vector of multiferroic BiFeO$_3$  at an ultrafast pace (namely of the order of picoseconds). We discuss the origin of such a double ultrafast switching, which is found to involve original intermediate magnetic states and may hold promise for designing various devices.
\end{abstract}

\maketitle

The magnetic chirality vector,  ${\bf \kappa}$, is  proportional to ${\bf S}_i\times {\bf S}_j$, where 
${\bf S}_i$ and ${\bf S}_j$ are spins  of two neighboring sites i and j, respectively.
This chirality vector is an important  quantity that is related to many interesting phenomena,  such as spin-polarized current, spin torque, 
anomalous Hall Effect, magnetic motors \cite{chir0,chir1,ch1,ahe},  etc.. It is extensively studied in the context of nano-magnetism \cite{ch2}.  Being able to  control and even switch this {\it magnetic} chirality vector by the application of an {\it electric} field can even further broaden the prospect of designing novel devices, in addition to be of obvious fundamental interest (by, e.g., understanding what governs this hypothetical magneto-electric coupling).  One may also wonder if such hypothetical switching can be ultrafast in nature, and what are the intermediate states involved in this switching (if any).

Here, we explore such possibilities in the multiferroic BiFeO$_3$ (BFO) system. We chose this material because (1) it possesses both a magnetic cycloid (which results in a non-zero ${\bf \kappa}$) and an electrical polarization at room temperature; and (2) previous studies have demonstrated that applying an electric field in BFO along a direction {\it being away} from the initial electric polarization leads to a rotation of this polarization, and, as a consequence of magneto-electric effects, to a change in the magnetic cycloidal plane \cite{zhao,cycloid0,lee}. It is thus interesting (and novel) to determine if applying an electric field that is {\it opposite} to the initial direction of the polarization can result in an {\it ultrafast} switching of both the polarization and the magnetic chirality vector, and how the dipoles and spins rearrange themselves during these possible latter switchings.
Via the use of  a first-principle-based method, we indeed discover that applying an electric field that is opposite to the initial direction of the polarization does indeed result in an ultrafast  (of the order of picosecond) switching of both the polarization and the magnetic chirality vector. Moreover, these  switchings involve striking intermediate states, which make them  promising for  applications and fundamentally interesting.

Technically, we use here the first-principles-based effective Hamiltonian method described in Ref.[\onlinecite{dovran}] for which the total energy is: 
\begin{equation}
E_{tot}=E_{FE-AFD}\biggl(\{{\bf u}_i\},\{{\bf \eta}_i\},\{{\bf \omega}_i\}\biggr)
+E_{Mag}\biggl(\{{\bf m}_i\},\{{\bf u}_i\},\{{\bf \eta}_i\},\{{\bf \omega}_i\}\biggr)
\end{equation}
where ${\bf u}_i$ is the local soft mode in the unit cell i, and is directly proportional to the electric dipole of that unit cell.  
$\{{\bf \eta}_i\}$ is the strain tensor and contains both homogeneous and inhomogeneous parts. 
The pseudo-vectors, ${\bf \omega}_i$'s, characterize the oxygen octahedra tilts, which are also termed antiferrodistortive (AFD) motions. More precisely, 
the direction of ${\bf \omega}_i$ is parallel to the axis about which the oxygen octahedron of the unit cell $i$ rotates, while the magnitude of ${\bf \omega}_i$ 
provides the angle of such rotation \cite{nonmag}. ${\bf m}_i$ represents the magnetic dipole moment of the Fe ion 
located inside the unit cell i, and has a fixed magnetic moment of 4$\mu_B$ -- as consistent with first principles \cite{mm}. 
The first term in the total energy (E$_{FE-AFD}$) gathers the energies associated with the structural degrees of 
freedom of the effective Hamiltonian (electric dipoles, strains, oxygen octahedra tilts) and their mutual interactions. It is described in Refs.[\onlinecite{Zhong,nonmag}].
The second term (E$_{Mag}$) is given in Ref.[\onlinecite{dovran}]. It
gathers the magnetic degrees of freedom, their mutual interactions and their couplings to the lattice degrees of freedom. Technically, it contains long-range magnetic dipolar interactions involving all Fe sites, short-range (up to third nearest neighbors) direct exchange interaction, and
magnetic exchange interaction mediated via the local modes, antiferro-distortive motions and the strains \cite{IgorBFO}. It also possesses an energy  that involves   oxygen octahedra tiltings and that 
yields a spin canting \cite{Al},  as well as another energetic term  that is derived from the spin-current model  \cite{Katsura,Aldo}. 
This latter has the analytical form:
\begin{equation}
E_{spin-current}=-C({\bf u}_i\times {\bf e}_{ij})\cdot({\bf m}_i\times{\bf m}_j)
\end{equation}
where ${\bf e}_{ij}$ is an unit vector in the direction joining 
second-nearest-neighbor  sites $i$ to $j$ and $C$ is a coefficient quantifying such interaction. We choose a value of $C$ equal to $6.3\times 10^{-6}$ Hartree/Bohr$\mu_B^2$, because 
it results \cite{dovran} in the well-known spin cycloid  of 
BFO bulk  \cite{cycloid0,cycloid1,cycloid2},  that propagates along a $<110>$ direction that is perpendicular to the direction of polarization.  The effective Hamiltonian is used inside molecular dynamics (MD) simulations that are 
solving the following equations of motions:\\
\begin{equation}
\begin{cases}
& M_{X}\frac{d^2X}{dt^2}=-\frac{\partial E_{tot}}{\partial X}\\
& \frac{d{\bf m}_i}{dt}=-\gamma{\bf m}_i\times[{\bf B}^i_{eff}(t)+{\bf h}^i(t)]-\gamma\frac{\lambda}{|{\bf m}_i|(1+\lambda^2)}{\bf m}_i\times\biggl({\bf m}_i 
\times[{\bf B}^i_{eff}(t)+{\bf h}^i(t)]\biggr)
\end{cases}
\end{equation}
where, in the first equation, $X$ is any (Voigt or Cartesian) component of the strain, local modes and oxygen octahedral tiltings, and $M_{X}$ represents the mass or moment of 
inertia associated with these degrees of freedom. 
This first equation therefore combines the effective Hamiltonian scheme with the conventional molecular dynamics simulations with variables ${\bf \eta}$, ${\bf u}_i$ and ${\bf \omega}_i$. 
The second equation is the stochastic (LLG) equation \cite{ll,gil}. 
There, the effective magnetic field ${\bf B}^i_{eff}$ acting on the magnetic dipole centered on the Fe site $i$ is 
given by ${\bf B}^i_{eff}=-\frac{\partial E_{tot}}{\partial{\bf m}_i}$. 
$\gamma$ is the gyro-magnetic ratio and $\lambda$ is the Gilbert damping constant that is  chosen here to be $\lambda=10^{-4}$. 
This particular choice for the damping constant is done because it was found to
satisfactorily  reproduce properties of BFO bulk \cite{dawei}. 
 ${\bf h}^i(t)$  is a fluctuating field  that allows to mimic finite-temperature effects. 
The Cartesian components of ${\bf h}^i(t)$ are associated with white noise processes represented by $<{\bf h}^i>=0$, and $<{\bf h}^i_\alpha(t)
{\bf h}^i_{\beta}(0)>=2\frac{\lambda k_BT}{\gamma |{\bf m}_i|}\delta_{\alpha,\beta}\delta(t)$, where $\alpha$ and $\beta$ are Cartesian components and where $\delta(t)$ is the
delta function and $\delta_{\alpha,\beta}$ is the Kronecker delta.

We use a $18\times 18\times 18$ supercell (29,160 atoms) that is periodic along the x-, y- and z-axes -- that are  lying along the pseudo-cubic [100], [010] and [001] directions, respectively. The dynamics are performed at 5K using NPT ensembles. 
The time step for  the MD simulation is 0.5 femtosecond (fs). 
As consistent with experiments \cite{cycloid2,r3c1,r3c3,r3c4}, 
the simulations predict a R3c ground state that is characterized by a polarization lying along the pseudo-cubic [111] direction and an antiphase-oxygen-octahedral-tilting vector,$<\omega>_R$,  
lying along this same [111] direction (note that  $<\omega>_R$ is defined as $\frac{1}{N}\sum_i\omega_i(-1)^{n_x(i)+n_y(i)+n_z(i)}$, where $n_x(i)$, $n_y(i)$ and $n_z(i)$ 
are the integers locating the site $i$ and $N$ is the number of Fe sites inside the supercell). The predicted magnetic ground state is also found to 
be consistent with measurements on BFO bulk  \cite{cycloid0,cycloid1,cycloid2,cheong1,cheong2}, since it is a spin cycloid propagating along 
the pseudo-cubic  [$\bar 101$] direction and for which third nearest neighbors along the $[111]$ polarization direction 
are aligned in an antiferromagnetic fashion. Such magnetic organization gives rise to 
a magnetic chirality vector (1) that is defined as  ${\bf \kappa}=\frac{1}{N_p}\sum_{i,j}{\bf m}_i\times {\bf m}_j$, where ${\bf m}_i$ and ${\bf m}_j$ 
are two neighboring spins along the propagation direction of the cycloid, and $N_p$ is the number of such pairs; and (2) that lies along the pseudo-cubic [$1\bar 21$] direction.

We now apply an electric field along the [$\bar{1}$$\bar{1}$$\bar{1}$] direction (that is opposite to the initial electrical polarization) and for which each Cartesian component has a magnitude of 7$\times$10$^8$ Volt/meter (V/m). Note that 
such field has the same order of magnitude than the one recently experimentally applied in BFO films, that is, 10$^9$ V/m \cite{Manunew}.
Let us first concentrate on the time evolution of the structural properties under this field.
Figure 1(a) shows the supercell average of the Cartesian components of the local modes, ($<u_x>$,$<u_y>$,$<u_z>$), as a function of time, while Figure 1b displays similar data but for the $<\omega>_R$ antiferrodistortive vector.  Figure 1a indicates that the electrical polarization first rapidly shrinks in magnitude with time, while still lying along [111], and then vanishes at a time $\simeq$ 200fs. Increasing further the time above this critical value leads to a  polarization now lying along the
[$\bar{1}$$\bar{1}$$\bar{1}$]  direction (i.e., parallel to the applied electric field)  that increases in magnitude until reaching a stable value for time at and above $\simeq$ 300fs. In other words, Fig. 1a reveals an {\it ultrafast} switching of polarization which is very much desired in ferroelectric capacitors used in devices 
such as Ferroelectric Random Access Memory (or for multi-state storage devices \cite{scott,bibes}) 
in order to write data at very fast scale. We also numerically found (by, e.g., performing Fourier transforms of the local modes \cite{Aaaron})  that the predicted polarization switching is continuous, as similar to the case of PbTiO$_3$ films \cite{PTOswitching}. In other words, it does not involve domain nucleation and movement\cite{Domain}. 
 
Interestingly, Fig. 1b reveals that any Cartesian component of $<\omega>_R$ first 
increases with time until reaching a maximum for times about which the polarization is the smallest in magnitude, and 
then decreases until slightly oscillating around a fixed value when further increasing the time. 
The opposite behavior of the magnitude of the polarization and oxygen octahedral 
tilting from zero to $\simeq$ 300fs reflects the well-know competitive behavior of 
these two structural degrees of freedom \cite{nonmag,comp1,comp2,comp3,comp4,comp5,comp6}. Moreover and unlike the polarization, 
$<\omega>_R$ does not switch in direction and thus remains oriented along the [111] direction
at any time. Such lack of change of direction of $<\omega>_R$, even when the polarization is reversed, originates from the fact that the coupling between oxygen octahedral tilting and polarization is quadratic in both the AFD and FE degrees of freedom \cite{nonmag}.
Figures 1a and 1b therefore reveal that the ultrafast switching of the polarization 
 follows a path starting from one R3c state (for which the polarization and $<\omega>_R$ are both along the pseudo-cubic
[111] direction) and ending in another R3c state (for which the polarization is now along  [$\bar{1}$$\bar{1}$$\bar{1}$]  while the oxygen octahedral still tilt  about [111]) via a R${\bar 3}$c state that is characterized by a vanishing polarization and an enhanced $<\omega>_R$ that remains oriented along [111]. As demonstrated and detailed in the Supplemental material, the coupling between polarization and oxygen octahedral tilting in BFO (and its effect on coercive field \cite{Maksymovych2012}) provides a natural explanation for the ultrafast and homogeneous nature of the switching of the polarization, since this latter follows a path that is well-defined by its strong coupling  with the oxygen octahedral tilting. Note that this R3c $\rightarrow$ R${\bar 3}$c $\rightarrow$ R3c path requires the application of an electric field along a $<111>$ direction, unlike what is typically done in most experiments -- for which the electric field lies along a $<001>$ direction (see, e.g., Ref. \cite{Maksymovych2012} and references therein).

Let us now concentrate on the consequence of the switching of the polarization on magnetic properties. For that, Figures 2a and 2b display the time evolution of the magnetic chirality vector and of the magnitude of the G-type antiferromagnetic (AFM) vector. 
This latter vector is denoted as $<{\bf L}>_G$ and is given by 
$<{\bf L}>_G=\frac{1}{N}\sum_i (-1)^{n_x(i)+n_y(i)+n_z(i)}~{\bf m}_i$. 
Moreover, Figures 3(a)-(e) provide snapshots of
the magnetic configuration along a given [$\bar 101$]  line joining Fe ions being second nearest neighbors of each other for five different specific times, respectively. 
Figures 2 and 3 reveal a quite complex and interesting electric-field-driven sequence of magnetic reorganization inside BFO bulk. 
For instance, during the time interval ranging between zero and $\simeq$ 200fs for which the polarization shrinks in magnitude along the pseudo-cubic [111] direction, the magnetic chirality vector remains merely unchanged. This result indicates that the initial magnetic cycloid does not respond to this ultrafast change of the polarization, as confirmed by comparing Figs 3a and 3b that corresponds to the initial time and  $t=100fs$, respectively.  On the other hand, between $\simeq$ 200fs  and $\simeq$ 400fs,  ${\bf \kappa}$ oscillates a lot and can adopt rather small magnitude while the G-type AFM vector gains some considerable strength. We interpret such evolution as indicating that the magnetic sublattice has ``noticed'', with a time delay, that the polarization can be quite small during the application of a reverse electric field, and therefore the spin cycloid tends to be destroyed in favor of an intermediate state possessing a significant G-type AFM character -- as consistent with the fact that the term $C({\bf u}_i\times {\bf e}_{ij})$ in front of ${\bf m}_i\times{\bf m}_j$ in Eq. (2) is now small. Such tendency is confirmed by looking at Fig. 3c that shows the organization of the magnetic dipoles at $t$=350 fs: one can see that adjacent (second-nearest-neighbor) dipoles are rather parallel to each other, as expected in a G-type AFM structure. Note that we numerically found 
that the G-type AFM vector can rotate when time varies  between $\simeq$ 200fs  and $\simeq$ 400fs, e.g. it can lie along [101], [0$\bar{1}$0] or [$\bar{1}$$\bar{1}$$\bar{1}$] pseudo-cubic direction. Moreover, for times comprised between  400fs and  1650fs (i.e., 1.65 ps), the magnetic chirality vector is quite stable with  an intermediate value of ${\bf \kappa} \simeq -2{\bf x} +4 {\bf y} -2 {\bf z}$ (that is equal to $-\frac{{\bf \kappa}_{Init}}{2}$, ${\bf\kappa}_{Init}$ being the initial chirality of the cycloid) and the strength of $<{\bf L}>_G$ has now decreased to a value ranging between 0.5 and 1 $\mu B$. Such features indicate that the resulting magnetic configuration is another intermediate state that is characterized by the {\it coexistence} of a significant  G-type antiferromagnetism altogether with a magnetic cycloid that occurs within the same plane than the initial magnetic cycloid but with a chirality being {\it opposite} to the initial one. 
Figure 3d provides a snapshot of such magnetic re-organization for $t$=1000fs. 
Interestingly, we numerically found that this peculiar intermediate magnetic state owes is existence to a combination of magnetic damping and magnitude of the electric field. More precisely and as shown in the Supplemental material, decreasing the value of the damping coefficient for the presently selected magnitude of the electric field tends to shrink the time window for which this magnetic intermediate state can occur. This is because the magnetic degrees of freedom can more easily follow the effective magnetic field resulting from the energy of Eq. (2), when damping is smaller. On the other hand, decreasing the magnitude of the electric field for our presently selected and realistic value of the damping coefficient results in  smaller magnitude of the local modes, and thus leads to smaller effective magnetic field. As a result, the magnetic degrees of freedom experience more problems to follow the dynamics of the local modes, and thus spend more time in this intermediate magnetic state. Finally, for times above 1650fs and as shown in Fig. 3e, 
the G-type AFM vector is nearly annihilated and the magnetic arrangement now consists of a spin cycloid having a chirality vector that is exactly opposite to the initial one (as consistent with the analytical form of Eq. (2) that indicates that reversing the electrical polarization should revert the cross products between the ${\bf m}_i$ and ${\bf m}_j$ vectors). In other words, the complete switching of the magnetic chirality has occurred around 1.65ps, and has therefore taken around 6 times longer than the switching of the polarization but  is still ultrafast in nature. 
Such ultrafast switching of chiral state is very promising and much more advantageous 
compared to devices where switching of magnetic state is achieved  by injecting current, since the latter process generates much more Joule 
heating. Moreover, the presently discovered electric-field-driven magnetic switching has involved quite a complex sequence of magnetic  arrangements, which can, in fact, be taken as an advantage over systems exhibiting both a polarization ({\bf P}) and a magnetization ({\bf M}). As a matter of fact, while these latter systems can, in theory, exhibit four different states -- namely   (P,M), (- P, M), (P,-M) and (-P,-M) -- only  two of them are easily accessible \cite{2state} due to large magneto-electric couplings \cite{scott}. 
Here, in contrast, {\it three} different well-defined states can be easily accessed when applying a field along [$\bar{1}$$\bar{1}$$\bar{1}$]. They are:
(${\bf P},{\bf \kappa}_{Init}$)  for time ranging between 0 and  200fs, ($-{\bf P},-\frac{{\bf \kappa}_{Init}}{2}$) for $t$ being in-between 400fs and 1650fs, and ($-{\bf P},-{\bf \kappa}_{Init}$) for times above. 
Interestingly, applying then a field along [111] from the final ($-{\bf P},-{\bf \kappa}_{Init}$) state was also numerically found (not shown here) to lead to the situation in which the polarization and magnetic chiralities are precisely {\it reversed} with respect to the case indicated in Figs. 1, 2 and 3. As a result, a {\it fourth} state appears, which is a mixed state characterized by (${\bf P},\frac{{\bf \kappa}_{Init}}{2}$), therefore rendering the possibility of making a four-state memory device  \cite{scott,bibes} possible
(see schematization in Fig. 4). Such memories could have all the combined advantages of ferroelectric random access memory (FeRAM )\cite{FeRAM} and magnetic random access memory \cite{MRAM} (MRAM). For instance,  one would be able to write the data using very low power (as in FeRAM) and
read data in a non-destructively way (as in MRAM, since magnetic switching does not require any movement of ions).  Note that changes in magnetic chirality may be practically observed
via Kerr, Hall or optical effects \cite{Kerr,Hall}.

Note that we chose to report here properties at 5K in order to get better statistics (i.e., less noise) for physical properties, especially those related to magnetism. However, we also numerically found that results similar to those depicted here can systematically occur at room temperature, especially if one uses slightly larger electric fields (to overcome the temperature-induced fluctuation of the electric dipoles) and slightly smaller damping constant (to allow the magnetic degrees of freedom to more easily follow the coupled polarization dynamics).

We hope that our study deepens the knowledge of multiferroics, and will encourage investigations
aimed at confirming and exploiting our results.
We are grateful to M. Bibes for stimulating discussions.
We thank ARO Grant No. W911NF-12-1-0085 for personnel support of S.B.
and of NSF Grant No. DMR-1066158 for personnel support of D.R.
Office of Basic Energy Sciences, under contract ER-46612,
and ONR Grants No. N00014-11-1-0384 and N00014-12-1-1034
are also acknowledged for discussions with
scientists sponsored by these grants.           

\newpage
\begin{figure}[h]
\includegraphics[width=135mm,height=135mm]{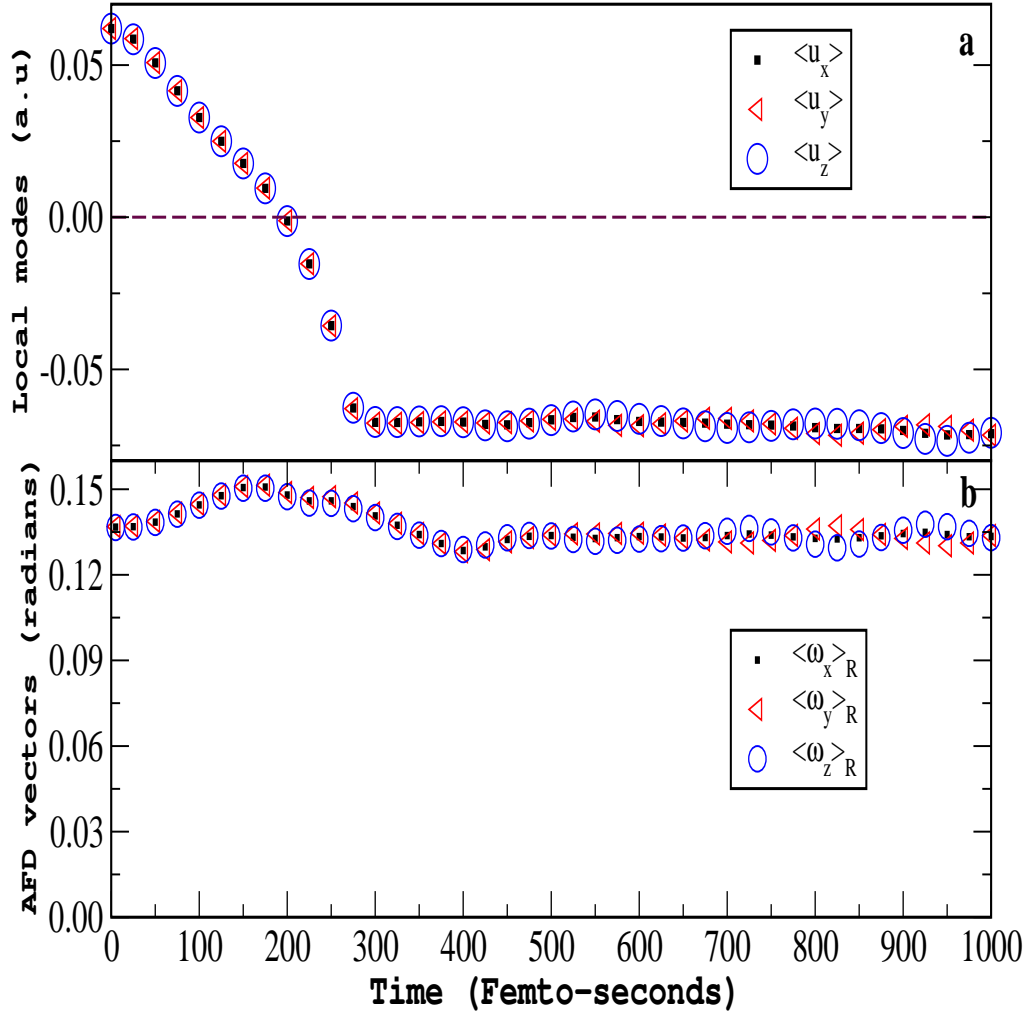}
\caption{(color online) Time evolution of the supercell average of the local modes (Panel a) and of the antiferrodistortive vector (Panel b)  in BFO bulks under a $dc$ electric field oriented along the [$\bar{1}$$\bar{1}$$\bar{1}$]  pseudo-cubic direction.}
\label{Fig.1}
\end{figure}
\begin{figure}[h]
\includegraphics[width=135mm,height=135mm]{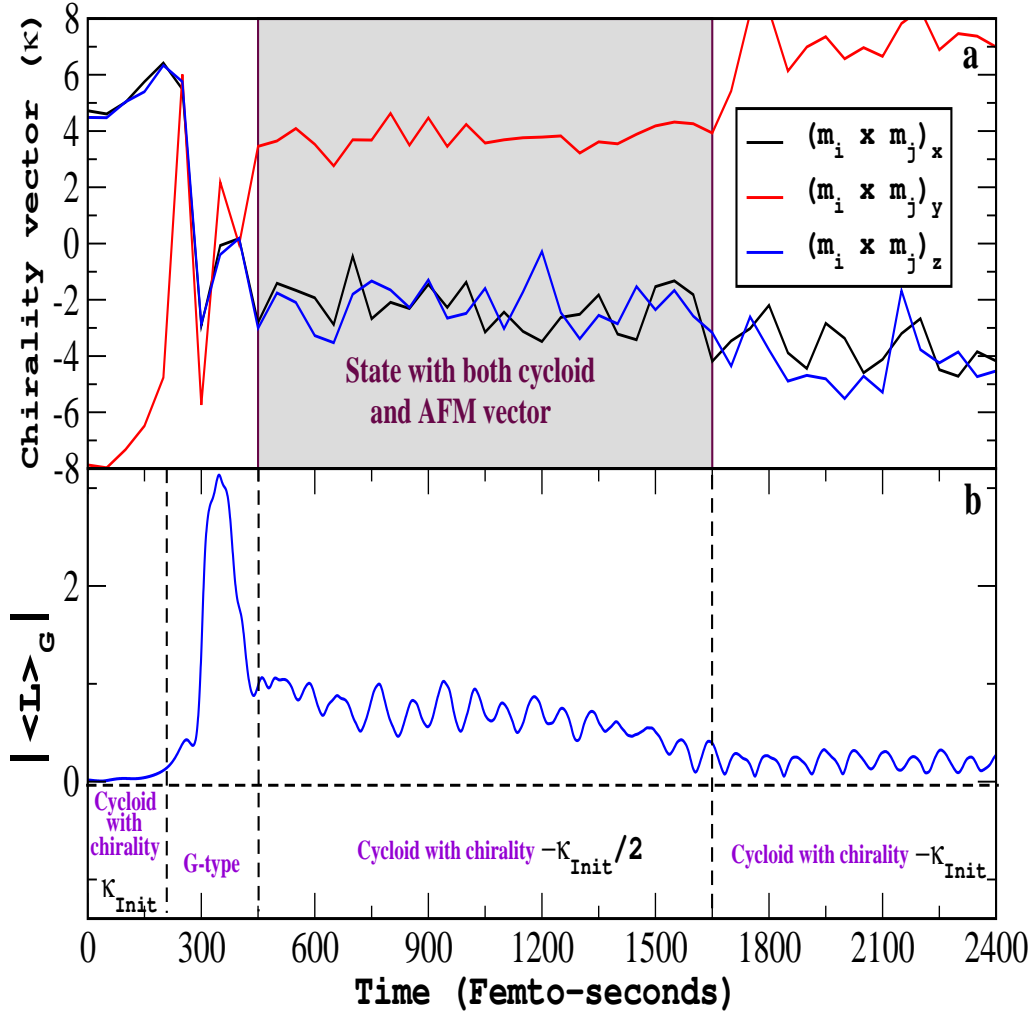}
\caption{(color online) Same as Figures 1 but for the magnetic chirality vector (Panel a) and the magnitude of the G-type AFM vector (Panel b).}
\label{Fig.2}
\end{figure}
\begin{figure}[h]
\includegraphics[width=135mm,height=145mm]{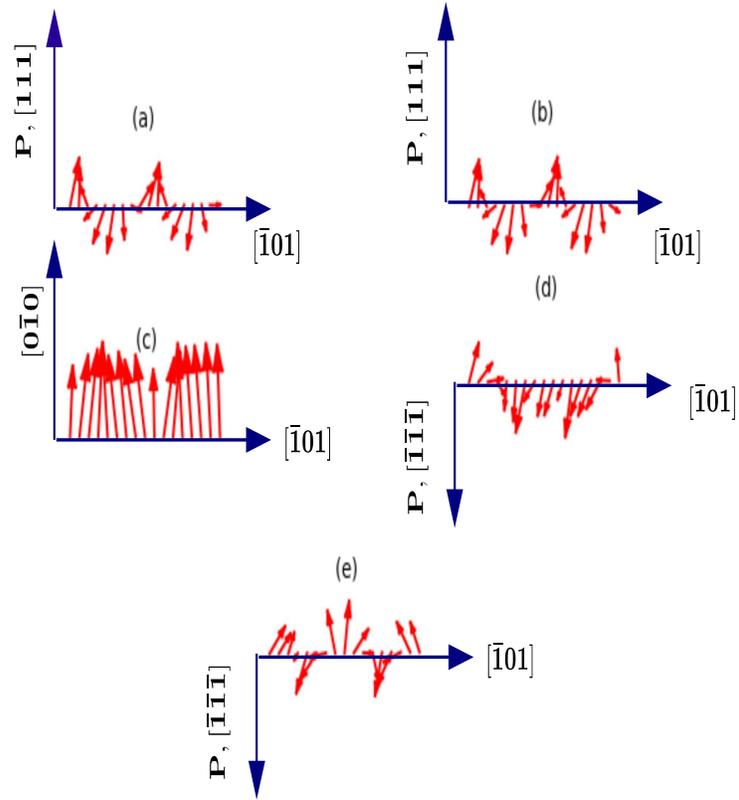}
\caption{(color online) Magnetic dipolar configurations along a specific [$\bar 101$]
line joining Fe ions for some specific times.
Panels (a), (b), (c), (d) and (e) correspond to a time of 0, 100, 350, 1000 and 3000 $fs$, respectively.}
\label{Fig.3}
\end{figure}
\begin{figure}
\includegraphics[width=175mm,height=140mm]{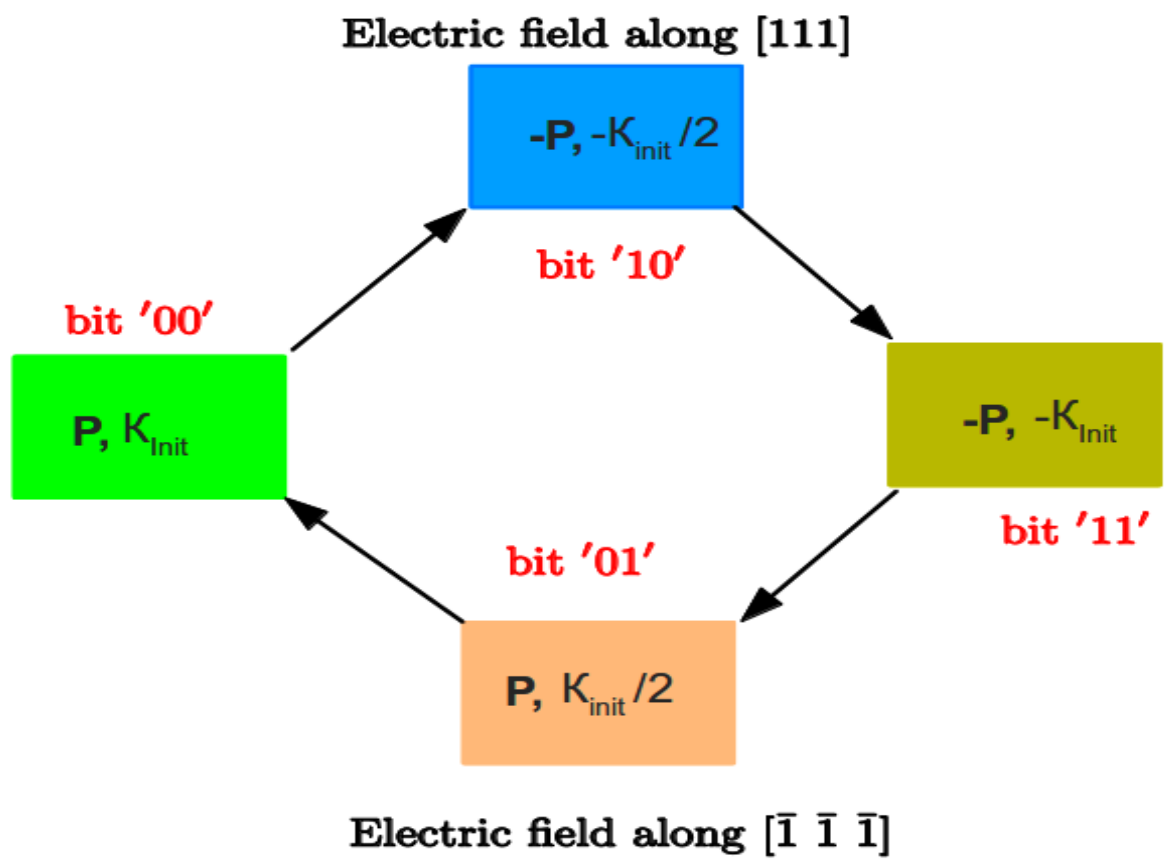}
\caption{(color online)  Schematization of the proposed four-state memory.}
\label{Fig.4}
\end{figure}

\end{document}


\title{Supplemental Material of \\ ``Ultrafast switching of the electric polarization and magnetic chirality \\ in BiFeO$_3$ by an electric field'' }
\author{S. Bhattacharjee$^{1}$, D. Rahmedov$^{1}$, Dawei Wang$^{2}$, Jorge \'I\~niguez$^{3}$  and L. Bellaiche$^{1}$}

\affiliation{$^{1}$Department of Physics and Institute for Nanoscience and Engineering, 
University of Arkansas\\ Fayetteville, Arkansas 72701, USA \\
    $^{2}$Electronic Materials Research Laboratory, Key Laboratory of the Ministry of Education and International Center for Dielectric
  Research, Xi'an Jiaotong University, Xi'an 710049, China \\
  $^{3}$Institut de Ci\`encia de Materials de Barcelona (ICMAB-CSIC), 
Campus UAB, 08193 Bellaterra, Spain}


\maketitle

The goals of this Supplemental Material are two-folds: (1) to provide details about the switching time and mechanism of the polarization; and (2) to give more information  about the intermediate magnetic state.

\section{I. Additional information about the polarization switching}

Figure 1a repeats Figure 1(a) of the manuscript, that is it displays the supercell average of the Cartesian components of the local modes, ($<u_x>$,$<u_y>$,$<u_z>$), as a function of time at the temperature T=5K, when BiFeO$_3$ (BFO) is under an electric field for which each Cartesian component has a magnitude of 7$\times$10$^8$ V/m and that is oriented along the [$\bar{1}$$\bar{1}$$\bar{1}$] pseudo-cubic direction. We will denote this case as Case 1.

Figure 1b shows the same data than Case 1, but with the important exception than the strength of the parameters representing the coupling of the electric dipoles with oxygen octahedral tiltings has been reduced by 25\% with respect to the case shown in Fig. 1a (i.e., these parameters are taken to be 75\% of their first-principles-derived values \cite{IgorBFO}). We will denote this case as Case 2. One can clearly see that the coupling between electric dipoles and oxygen octahedral tilting has a dramatic effect on the switching time since this latter is increased from $\simeq$ 300fs to $\simeq$ 900fs from Case 1 to Case 2.

Moreover, Figure 2a and 2b report the Fourier transform of the Cartesian components of the local modes \cite{Aaaron} at the zone-center of the 5-atom cubic first Brillouin zone ($\Gamma$ point) as a function of time for Case 1 and Case 2, respectively, still at T=5K. Note that a value of 1 for these Fourier transforms
characterize an {\it homogeneous} pattern of the local dipoles, independently of their magnitude \cite{Aaaron}. On the other hand, deviation from 1 implies that other k-points are involved (since the sum over the k-points of the Fourier transform of any component of the local modes should be equal to unity \cite{Aaaron}) and thus that the electric dipole pattern is {\it inhomogeneous}.
 For Case 1, such Fourier transforms are always very close to their maximum values of 1, revealing that the polarization switching is {\it homogeneous} (the local electric dipoles are all along the [111] direction, and simply change their magnitude and then sign as time evolves).
On the other hand, for Case 2, these Fourier transforms all significantly deviate from 1 for times $t$ ranging between  $\simeq$ 450fs and $\simeq$ 900fs, and even nearly vanish when $t$ $\simeq$ 600fs
(for which the macroscopic polarization is nearly null, see Fig. 1(b)). Such behaviors point towards an {\it inhomogeneous} switching mechanism of the polarization, when the coupling between local electric dipoles and oxygen octahedral tiltings has been reduced with respect to its  first-principle value in BFO.

Moreover, Figure 3a reports snapshots of the pattern of the local electric dipoles for different characteristic times for Case 1. Figure 3b shows similar data, but for Case 2. The different character of the switching mechanism of the polarization (i.e., homogeneous for large coupling between electric dipoles and oxygen octahedral tiltings {\it versus} inhomogeneous for smaller coupling between electric dipoles and oxygen octahedral tiltings ) is further evidenced there, when concentrating on times for which the macroscopic polarization is nearly zero (i.e, $t$ $\simeq$ 200fs for Case 1 {\it versus} $t$ $\simeq$ 600fs for Case 2).

Let us also emphasize that the {\it direction} of the electric field also matters  for the nature of the switching of the polarization. For that, let us consider again that the coupling between local modes and oxygen octahedral tiltings is the one provided by first-principles (as in Case 1), but  we rotate the direction of the applied electric field (without changing its magnitude). Such change defines Case 3. More precisely, Case 3 corresponds to an electric field having $-8$$\times$10$^8$ V/m, $-8$$\times$10$^8$ V/m and $-4$$\times$10$^8$ V/m for its $x$, $y$ and $z$-Cartesian components.
Figures 4a and 4b  show the time-dependency of  ($<u_x>$,$<u_y>$,$<u_z>$) and of the
Fourier transform of the Cartesian components of the local modes  at the $\Gamma$ point, respectively,
at T=5K for Case 3. Comparing such figures with those corresponding to Case 1 (i.e., Figs. 1(a) and 2(a)) tells us that the main effect of deviating the direction of the applied electric field is to render the switching of the polarization more inhomogeneous (as evidenced by the reduction of the Fourier transforms shown in Fig. 4b between 150 and 300 fs). Such results can be understood by realizing that the coupling between electric dipoles and oxygen octahedral tiltings naturally 
forces the polarization to be along the same direction than the axis about which the oxygen octahedral tilt \cite{BFOPTIgor}, and thus applying an electric field deviating from such axis results in an effective decrease of the role that oxygen octahedral tilts play on polarization.

In order to further understand the switching mechanism of the polarization, Figure 5 shows 
the  T=5K equilibrated ($<u_x>$,$<u_y>$,$<u_z>$) Cartesian components of the local modes as a function of an {\it increasing} magnitude of a $dc$ electric field applied along   [$\bar{1}$$\bar{1}$$\bar{1}$] for two different cases: one for which the  coupling between local modes and oxygen octahedral tiltings is the one provided by first-principles {\it versus} another case for which this coupling is reduced by 25\%. Practically, these equilibrated Cartesian components are those adopted by the system after a rather long time (namely, 13,000fs) has passed during the molecular dynamics simulations. Note that the set-up of these simulations differs from those associated with 
Figures 1 and 2, in these sense that, for these latter, a large electric field was suddenly applied 
along [$\bar{1}$$\bar{1}$$\bar{1}$]   to the system (and kept frozen), and the resulting behavior of physical properties was then monitored as a function of time. 
In contrast, for Fig. 5 the field is progressively increased in magnitude, and the polarization equilibrated at 5K is recorded for each considered magnitude of the electric field. By proceeding in this way, we avoid injecting a big {\em excess energy} in our simulated system, and we can thus follow {\em quasi-adiabatically} the evolution of the $[111]$-polarized ferroelectric state up to its (meta)stability limit, which corresponds with the theoretical value of the thermodynamic coercive field, $E_c$. Beyond that field value, the system finally transforms into the switched configuration.
Figure 5 reveals that decreasing the strength of the coupling between the local modes and the oxygen octahedral tilting shifts the  coercive field to larger value. For instance, each Cartesian component of $E_c$ is slightly smaller in magnitude than $7$$\times$10$^8$ V/m when the coupling is taken to be the one provided by first principles while it is as high as $8.65$$\times$10$^8$ V/m when this coupling has been reduced by 25\%.
Interestingly, taking into account the magnitude of this coercive field and its dependency on the coupling between local modes and oxygen octahedral tiltings  can also lead to an alternative (complementary) explanation of the character of the switching mechanism of the polarization. More precisely,  an homogeneous (respectively, inhomogenous) switching of the polarization can occur when the electric field that is suddenly applied to the system is larger (respectively, smaller) in magnitude than $E_c$ -- which was the situation for Case 1 (respectively, Case 2).
To further confirm such possibility, we performed additional calculations that only differ from  Case 2 by the magnitude of the electric field suddenly applied along [$\bar{1}$$\bar{1}$$\bar{1}$] : it is now such as each of the Cartesian component is equal to -$9$$\times$10$^8$ V/m, which thus renders this applied field larger in magnitude than $E_c$ when the coupling  between the local modes and the oxygen octahedral tilting is taken to be equal to 75\% of its first principle value (see the curve drawn in blue in Figure 5). In that new case, we found (not shown here) that the switching of the electrical polarization becomes homogeneous, unlike in Case 2, and much faster than in Case 2 (namely, around 300fs).  Such results further demonstrate the importance of comparing the magnitude of the applied field with the coercive field when trying to understand the nature and time of the switching mechanism of the polarization.

\section{II. Additional information about the Intermediate magnetic state}

Table I reports the life time of the intermediate magnetic state (denoted as  $-\frac{{\bf \kappa}_{Init}}{2}$ in the manuscript) for different chosen values of the damping constant ($\lambda$) at T=5K, when BFO is under an electric field lying along the [$\bar{1}$$\bar{1}$$\bar{1}$] direction and for which each Cartesian component has a magnitude of 7$\times$10$^8$ V/m. Typically, increasing the damping coefficient enhances the life time of this intermediate state.

We also numerically found (not shown here) that the transition from the $-\frac{{\bf \kappa}_{Init}}{2}$ state to the $-{\bf \kappa}_{Init}$ chiral state can be more continuous rather than sharp, if one slightly increases the damping coefficient with respect to the one used for obtaining the results shown in Fig. 2a of the manuscript. All these findings are consistent with the effects of the damping 
constant on the switching time and the discontinuous nature of the switching mechanism at intermediate value of the damping constant  discussed  in Ref. \cite{Kikuchi}. 
 
 Furthermore, Table II shows how this life-time now depends on the magnitude of the applied field (that is still oriented along  [$\bar{1}$$\bar{1}$$\bar{1}$]), for a fixed (realistic) damping coefficient $\lambda=10^{-4}$ at a temperature of 5K. Interestingly, decreasing this magnitude leads to a larger life-time of the $-\frac{{\bf \kappa}_{Init}}{2}$ magnetic state -- which can even be completely stabilized for electric field having all its Cartesian components equal to, or smaller than, 6 $\times$10$^8$ V/m. Such features are promising for stabilizing or refreshing (as in DRAM) the intermediate state by a time-dependent electric field (that will have each of its Cartesian component ranging, e.g.,  between $6 \times 10^8$ and $7 \times 10^8$ V/m).
    
Finally, let us point out that we also numerically found (not shown here) that  starting from the $-\frac{{\bf \kappa}_{Init}}{2}$ state  and applying an electric field that is oriented along  [111] and that has any of its Cartesian component equal to $5.5 \times 10^8$ V/m can lead to a direct switching
from  $-\frac{{\bf \kappa}_{Init}}{2}$ to a chiral state that is very close to  $+\frac{{\bf \kappa}_{Init}}{2}$. 
\begin{table}
\begin{tabular}{l|c}
\hline
Damping constant & Life-time of the intermediate\\
$\lambda$ & state (in femtoseconds)\\
\hline
$10^{-4}$ & 1200 \\
$8\times 10^{-5}$ & 700 \\
$6\times 10^{-5}$ & 900 \\
$4\times 10^{-5}$ & 400 \\
$2\times 10^{-5}$ & $\sim$~ 400\\
$10^{-5}$ & $\sim$~ 100\\
\hline
\end{tabular}
\caption{Dependence of the life-time of the intermediate state with respect to magnetic damping when an electric field, having each of its Cartesian component equal to  
$7\times 10^8$ V/m in magnitude, is applied along the [$\bar{1}$$\bar{1}$$\bar{1}$] direction at 5K.}
\end{table}
\begin{table}
\begin{tabular}{l|c}
\hline
Cartesian component of the Electric field & Life-time of the intermediate\\
(in V/m) & state (in femtoseconds)\\
\hline
$7.5\times 10^8$ & 900 \\
$7\times 10^8$ & 1200 \\
$6.5\times 10^8$ & 1800 \\
$6\times 10^8$ & $\infty$ \\
\hline
\end{tabular}
\caption{Dependence of the life-time of the intermediate state on the magnitude of an electric field applied along [$\bar{1}$$\bar{1}$$\bar{1}$] at 5K, for a fixed damping constant of $10^{-4}$.}
\end{table}

\newpage
\begin{figure}[h]
\includegraphics[width=130mm,height=135mm]{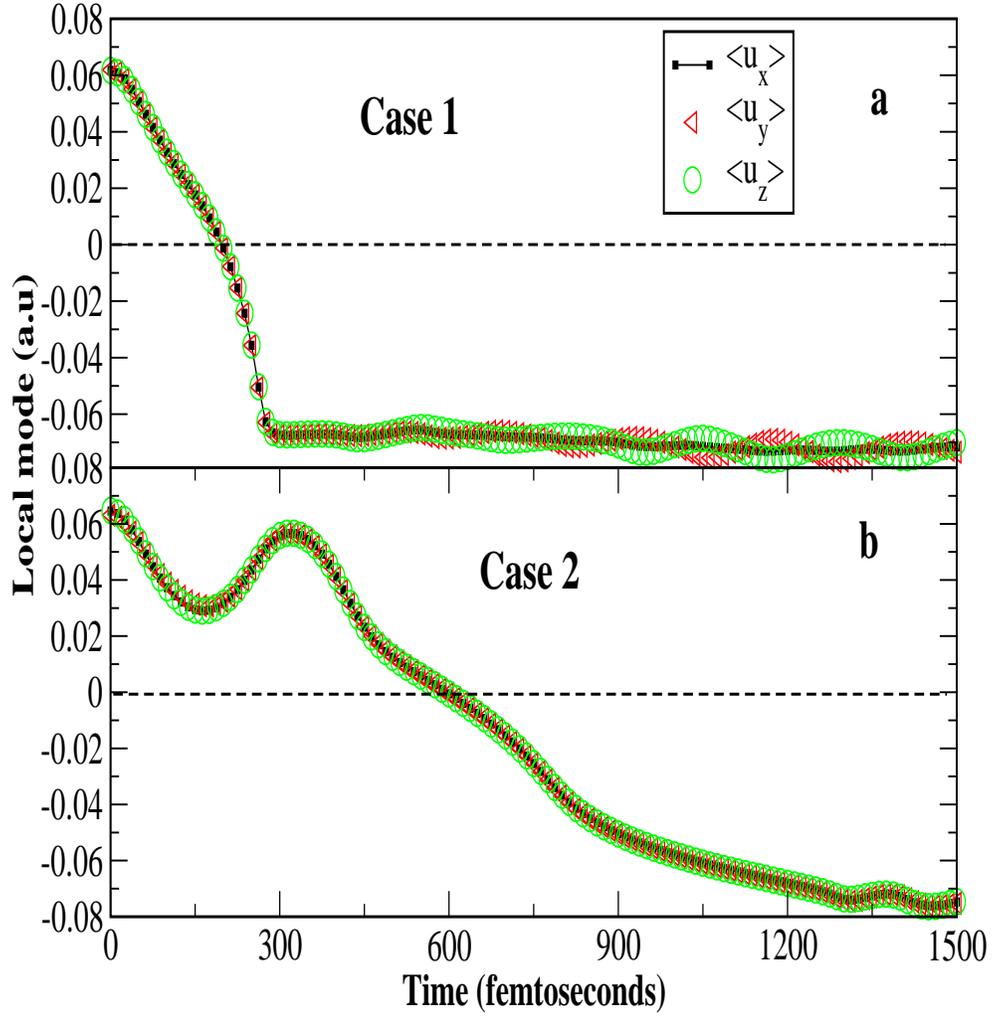}
\caption{(color online) Time evolution of the supercell average of the local modes for Case 1(Panel a)  and Case 2 (Panel b) at 5K.}
\label{Fig.1}
\end{figure}
\begin{figure}[h]
\includegraphics[width=130mm,height=135mm]{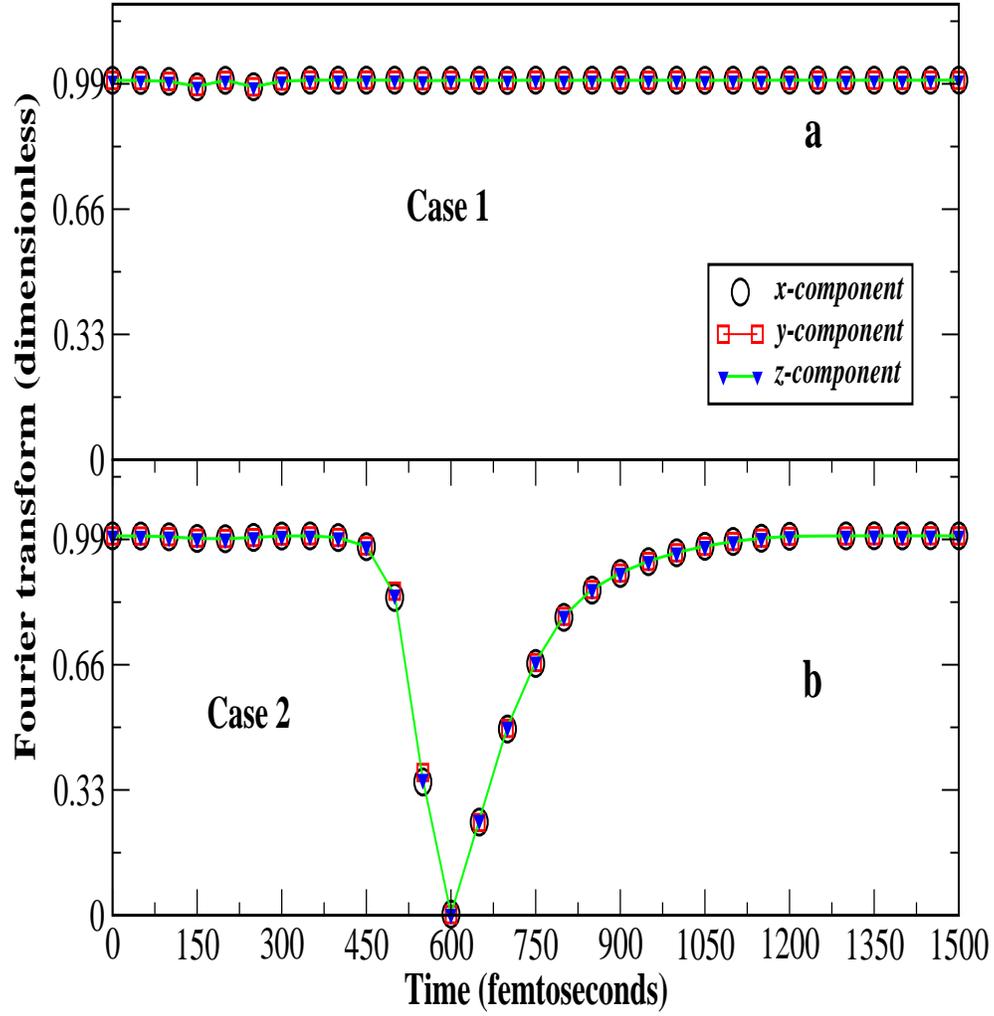}
\caption{(color online) Time evolution of the Fourier transform of the Cartesian components of the local modes at the $\Gamma$-point for Case 1 (Panel a) and Case 2 (Panel b)  at 5K.}
\label{Fig.2}
\end{figure}
\begin{figure}[h]
\includegraphics[width=140mm,height=145mm]{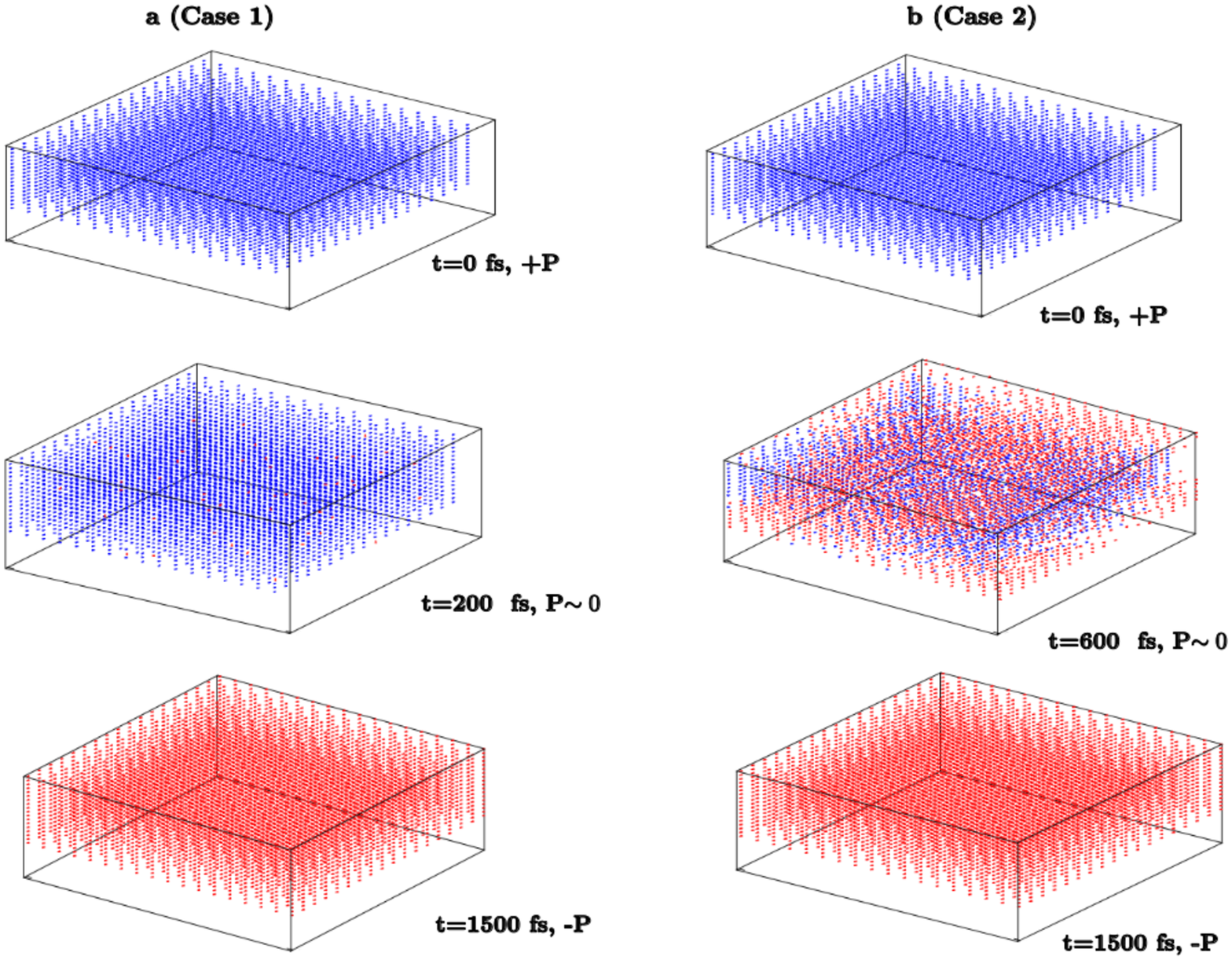}
\caption{(color online) Snapshots of the time evolution of the local electric dipoles for different characteristic times for Case 1 (Panel a) and Case 2(panel b)  at 5K. The blue and red colors indicate electric dipoles having positive and negative z-component, respectively. As a result, the obvious mixing of blue and red arrows in the middle inset of Panel (b) clearly shows the inhomogeneous nature of the switching of the polarization in Case 2. On the other hand, the switching in Case 1 occurs via the reduction of the 
magnitude of the dipoles (see decrease of the length of the blue arrows from top to middle insets in Panel (a)) and then a reversal of the Cartesian components of the electric dipoles (see the change from  a majority number of blue arrows to red arrows when going from the middle to bottom inset of Panel (a)).}
\label{Fig.3}
\end{figure}
\begin{figure}[h]
\includegraphics[width=130mm,height=135mm]{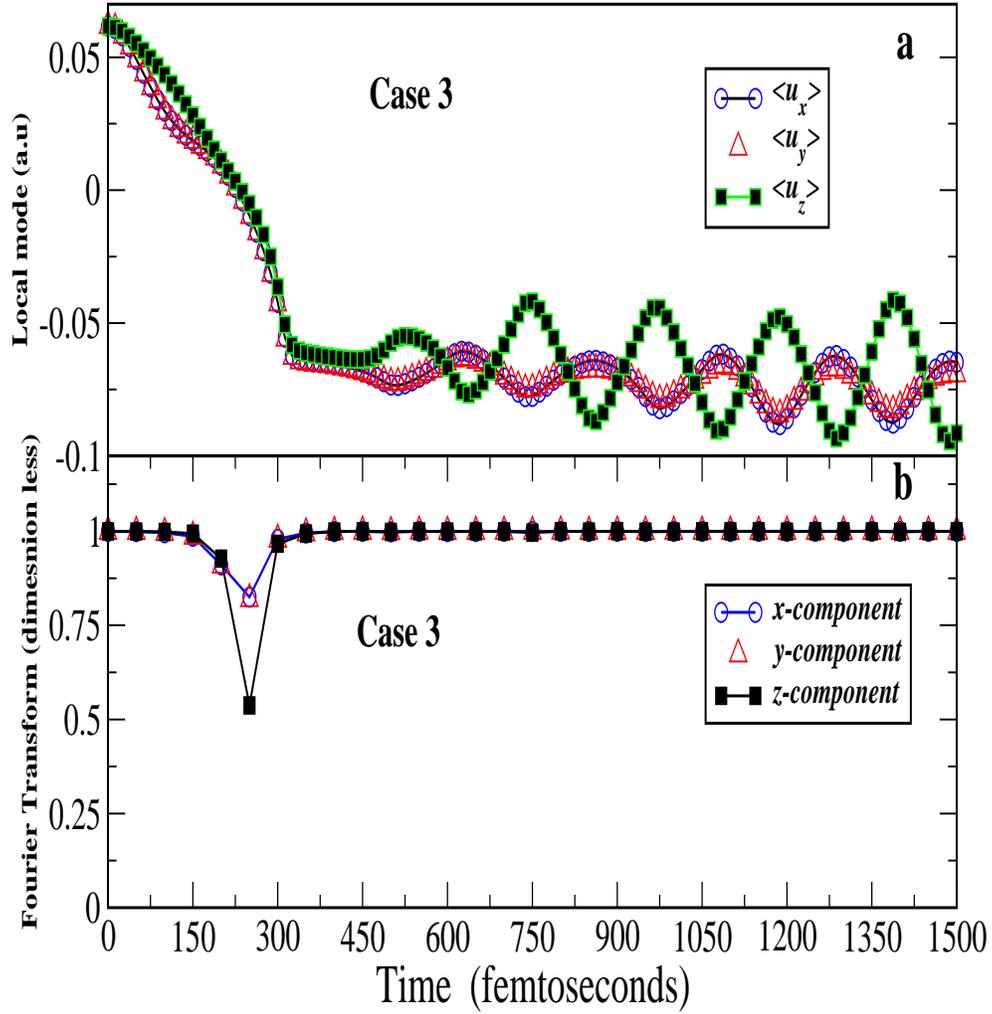}
\caption{(color online) Time evolution of of the supercell average of the local modes (Panel a) and of the Fourier transform of the Cartesian components of the local modes 
at the $\Gamma$-point (Panel b) for Case 3  at 5K.}
\label{Fig.4}
\end{figure}
\begin{figure}[h]
\includegraphics[width=130mm,height=135mm]{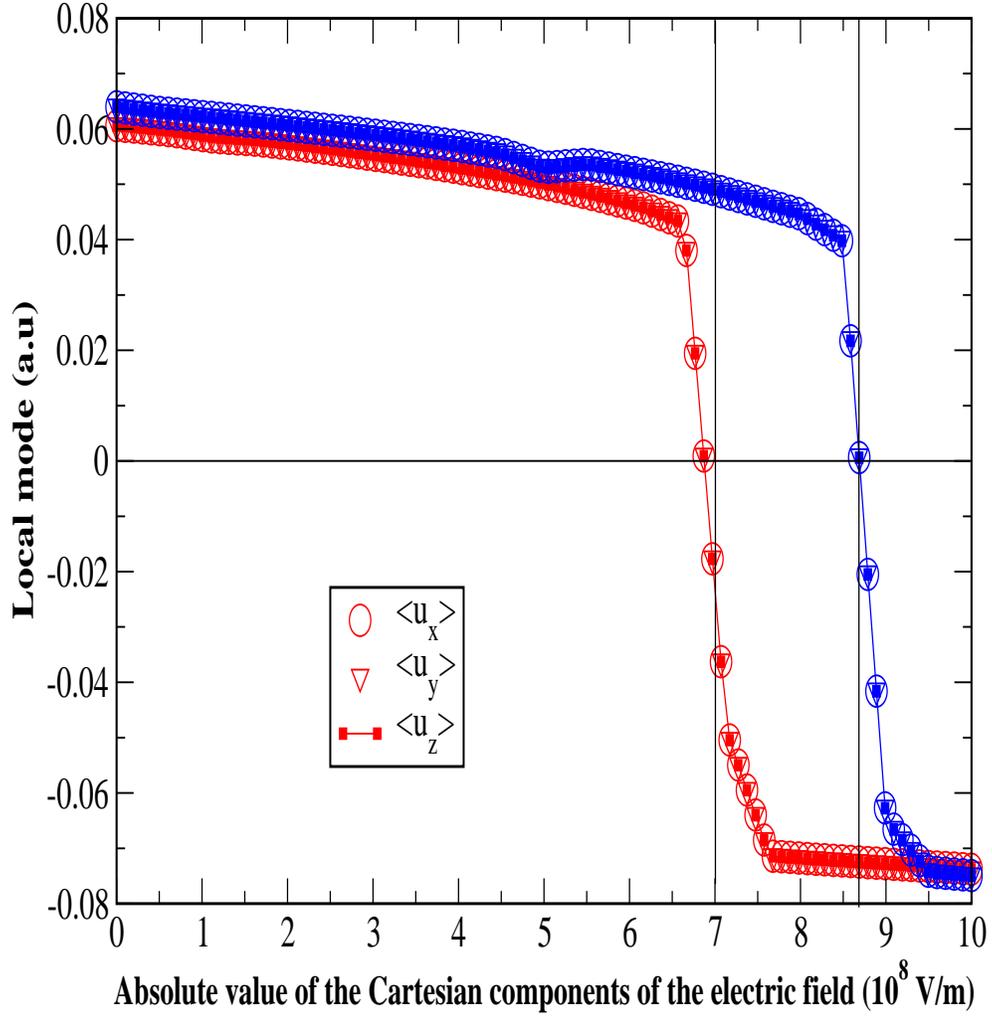}
\caption{(color online) Dependence of the equilibrated supercell average of the local modes as a function of the absolute value of the Cartesian components of an electric field applied along
[$\bar{1}$$\bar{1}$$\bar{1}$] at T=5K.  Symbols in red correspond to a coupling between local modes and oxygen octahedral tiltings being provided by density-functional calculations, while symbols in blue depict results for this latter coupling being reduced by 25\%.}
\label{Fig.5}
\end{figure}